\documentclass[twocolumn,amsmath,amssymb,showpacs,prl,superscriptaddress]{revtex4-2} 

\bibstyle{apsrev}
\usepackage{bm}%
\usepackage{graphicx}
\usepackage{tikz}
\usepackage{color}
\usepackage{xcolor}
\usepackage{physics}
\usepackage[colorlinks=true,linkcolor=blue]{hyperref}%

\expandafter\ifx\csname package@font\endcsname\relax\else
 \expandafter\expandafter
 \expandafter\usepackage
 \expandafter\expandafter
 \expandafter{\csname package@font\endcsname}%
\fi
\begin{document}

\definecolor{pyblue}{HTML}{1F77B4}
\definecolor{pyorange}{HTML}{FF7F0C}
\definecolor{pygreen}{HTML}{2CA02C}
\definecolor{pyred}{HTML}{D62728}
\definecolor{jlblue}{rgb}{0.0,0.6056031611752245,0.9786801175696073}
\definecolor{jlorange}{rgb}{0.8888735002725198,0.43564919034818994,0.2781229361419438}
\newcommand*\Diff[1]{\mathop{}\!\mathrm{d^#1}}

\title{Controlling the dewetting morphologies of thin liquid films by switchable substrates}

\author{S. Zitz}
\email{s.zitz@fz-juelich.de}
 \affiliation{Helmholtz Institute Erlangen-N\"urnberg for Renewable Energy,\\
  Forschungszentrum J\"ulich,
  F\"urther Strasse 248, 90429 N\"urnberg, Germany}%
  \affiliation{Department of Chemical and Biological Engineering, Friedrich-Alexander-Universit\"at Erlangen-N\"urnberg, F\"{u}rther Stra{\ss}e 248, 90429 N\"{u}rnberg, Germany}
\author{A. Scagliarini}%
\email{andrea.scagliarini@cnr.it}
 \affiliation{Institute for Applied Mathematics "M. Picone" (IAC), 
Consiglio Nazionale delle Ricerche (CNR),\\
Via dei Taurini 19, 00185 Rome, Italy}%
\affiliation{INFN, sezione Roma ``Tor Vergata'', via della Ricerca Scientifica 1, 00133 Rome, Italy}
\author{J. Harting}
\email{j.harting@fz-juelich.de}
 \affiliation{Helmholtz Institute Erlangen-N\"urnberg for Renewable Energy,\\
  Forschungszentrum J\"ulich,
  F\"urther Strasse 248, 90429 N\"urnberg, Germany}%
 \affiliation{Department of Chemical and Biological Engineering and Department of Physics, Friedrich-Alexander-Universit\"at Erlangen-N\"urnberg, F\"{u}rther Stra{\ss}e 248, 90429 N\"{u}rnberg, Germany}
\date{\today}

\begin{abstract}
\noindent Switchable and adaptive substrates emerged as valuable tools for the control of wetting and actuation of droplet motion. Here we report a computational study of the dynamics of an unstable thin liquid film deposited on a switchable substrate, modelled with a space and time varying contact angle.
With a static pattern, all the fluid is drained into droplets located around contact angle minima, whereas for a sufficiently large rate of wettability variation a state consisting of metastable rivulets is observed. 
A criterion discriminating whether rivulets can be observed or not is identified in terms of a single dimensionless parameter.
Finally, we show and explain theoretically how the film rupture times, droplet shape and rivulet life time depend on the pattern wavelength and speed.  
\end{abstract}

\maketitle

\newcommand{\ts}{\textsuperscript}

\noindent {\it Introduction}. Wet surfaces and droplets are part of our every-day experience and of numerous industrial processes including coating, tribology, painting and printing, to name but a few~\cite{gross1980fluid,szeri2010fluid,DERYCK1998278,doi:10.1146/annurev.fluid.31.1.347,DASILVASOBRINHO19991204,singh2010inkjet,jo2009evaluation,Wijshoff2010}. 
Moreover, the continuously growing technological interest for lab-on-a-chip devices~\cite{C6LC00387G,Focke} as well as for printable electronics~\cite{Kim_2005, Luechinger_2008} and printable photovoltaics~\cite{Brabec2001,RH20}, whose efficiency relies crucially on a precise control of material deposition upon (de-)wetting of liquid films, drew the attention to applications where the substrate is adaptive or switchable, i.e. it is not inert but responds dynamically to external stimuli or to the evolution of the coating liquid film itself~\cite{ButtEtAl_Langmuir2018,GuoGuo_RSCAdv2016}. 
Several realizations of switchable and adaptive substrates have been proposed~\cite{LiuChem2005, XiCSR2010}, involving smart materials such as polymer brushes~\cite{CohenStuartEtAl_NatMat2010,ayres2007stimuli,YongEtAl_Mat2018}, thermal-responsive hydrogels~\cite{ChenEtAl_SM2010}, light-responsive molecules and microstructures~\cite{IchimuraEtAl_Science2000,DelormeEtAl_Langmuir2005,OscuratoEtAl_AAMI2017} or processes such as electrowetting~\cite{MugeleEtAl_JPCM2005}.
Probably the simplest, yet non-trivial, modelling of a switchable substrate can be realized by a space and time dependent wettability pattern~\cite{GrawitterStark1}. 
While a consistent body of theoretical/computational work has been devoted to processes on static heterogeneous substrates, though, the time dependent case is still almost unexplored, with few relevant exceptions focusing on single droplet spreading and sliding~\cite{GrawitterStark1,GrawitterStark2,ThieleHartmann} or limited to an analysis of the linear regime~\cite{suman2006dynamics}.

In this Letter, we study, by means of numerical simulations, the full dewetting dynamics of a thin liquid film deposited over a substrate with a time varying wettability pattern, from rupture to the long time morphology. 
We identify two regimes where the rupture times grow with the pattern wavelength either linearly (on the static pattern) or attain a constant value (in the time dependent case), for short wavelengths, and approach a quadratic law as the wavelength increases. These observations are then explained theoretically.
We show that, by tuning the rate of change (the ``speed'') of the underlying pattern, it is possible to control, to some extent, the dewetting morphology. In particular, for large enough pattern speeds, we detect a metastable state, where the film retracts into metastable rivulets, eventually breaking up into multiple droplets.
We introduce a control parameter that discriminates whether rivulets or just droplets (as in the static situation) can be observed and we propose a phenomenological argument that justifies the logarithmic dependence of the rivulets life time on the pattern speed.\\ 
\\
\noindent {\it Method.} In order to simulate the dewetting dynamics on patterned, ``switchable'', substrates, we integrate numerically the thin-film equation~\cite{ReynoldsLubr,RevModPhys.69.931,RevModPhys.81.1131} 
\begin{equation}\label{eq:thinfilm}
    \partial_t h(\mathbf{x},t) = \nabla\cdot\left(M_{\delta}(h)\nabla p(\mathbf{x},t)\right),
\end{equation}
by means of a recently developed method, based on a lattice Boltzmann (LB) scheme~\cite{PhysRevE.100.033313,PhysRevE.104.034801}.
Eq.~(\ref{eq:thinfilm}) describes, in a lubrication approximation spirit, the evolution of the height field (film thickness) $h(\mathbf{x},t)$, denoting the location of the liquid/air interface. The mobility function 
\begin{equation}\label{eq:mobility}
  M_{\delta}(h) = \frac{2h^3 + 6\,\delta\, h^2 + 3\delta^2h}{6\mu},
\end{equation}  
depends on the velocity boundary condition at the substrate, parameterized by an effective slip length $\delta$
(for $\delta \rightarrow 0$ it reduces to the no-slip form $h^3/(3\mu)$). Here, $\mu$ is the fluid dynamic viscosity.
The film pressure $p(\mathbf{x},t)$ consists of the sum of the Laplace and disjoining pressures, that is $p(\mathbf{x},t) = -\gamma \nabla^2 h - \Pi$. 
The disjoining pressure $\Pi$ can be seen as (minus) the derivative, with respect to the film thickness, of an effective interfacial potential. As such, it contains the information on the liquid/solid and solid/gas interactions and, hence, on the wettability, which is parameterized in terms of the contact angle $\theta$~\cite{RevModPhys.81.739, SCHWARTZ1998173}. 
The expression adopted for $\Pi$ is
\begin{equation}\label{eq:disjoinpressure}
  \Pi(h,\theta) = \frac{2\gamma}{h_{\ast}}(1-\cos(\theta(\mathbf{x},t)))\left[\left(\frac{h_{\ast}}{h}\right)^3 - \left(\frac{h_{\ast}}{h}\right)^2\right].
\end{equation}
$h_{\ast}$ is the height at which the disjoining pressure vanishes and it sets the precursor layer thickness.
The time variability of the patterned substrate then enters the model precisely through the disjoining pressure, by making the contact angle space and time dependent, i.e. $\theta = \theta(\mathbf{x},t)$.
In particular, we employ the sinusoidal form
\begin{equation}\label{eq:sinetheta}
   \!\! \theta(\mathbf{x},t) = \theta_0 + \delta\theta\left[\sin\left(q_{\theta} (x+v_{\theta x}t)\right)\sin\left(q_{\theta}(y+v_{\theta y}t)\right)\right],\! 
\end{equation}
where $q_{\theta} = 2\pi/\lambda$, i.e.~the pattern evolves in time as a plane wave. 
We fix the velocity direction to one diagonal, namely $\mathbf{v}_{\theta} = (v_{\theta x},v_{\theta y}) = v_{\theta}(1/\sqrt{2},-1/\sqrt{2})$ (we will return later on the importance of this choice), and we set $\theta_0 = 20^{\circ}$ and $\delta\theta=10^{\circ}$~\footnote{Notice that, since a typical velocity is such that $v_{\theta} \Delta t \ll \Delta x$ (in one LB time step $\Delta t$ the wave would travel a distance much smaller than a lattice spacing $\Delta x$),
the time update needs to be interpreted in an integer part sense, that is the pattern is shifted by one $\Delta x$ every $1/v_{\theta x}$ time steps (and equivalently in the $y$-direction).}.
Hereafter, lengths will be expressed in units of the mean film height, $h_0$ (which is constant in time, due to mass conservation), whereas the characteristic time and velocity scales are given by
\begin{equation}\label{eq:t0}
t_0 = \frac{3\mu}{\gamma h_0^3 q_0^4}, \quad v_0 = \frac{\lambda}{t_0},
\end{equation}
where $\gamma$ is the surface tension (the numerical values, in LB units, are set to $\gamma = 0.01$ and $\mu=1/6$).
The thus defined $t_0$ is the inverse growth rate of the most unstable mode (of wavenumber $q_0$) of a spinodally dewetting film (on a homogeneous substrate) \cite{Mecke_2005,PhysRevE.100.023108}. In our heterogeneous case, we define the wavenumber $q_0$ as $q_0^2=\frac{1}{2\gamma}\frac{\partial \Pi(h,\theta)}{\partial h}(h=h_0,\theta=\theta_0+\delta \theta)$.
\begin{figure}
    \centering
    \includegraphics[width=0.45\textwidth]{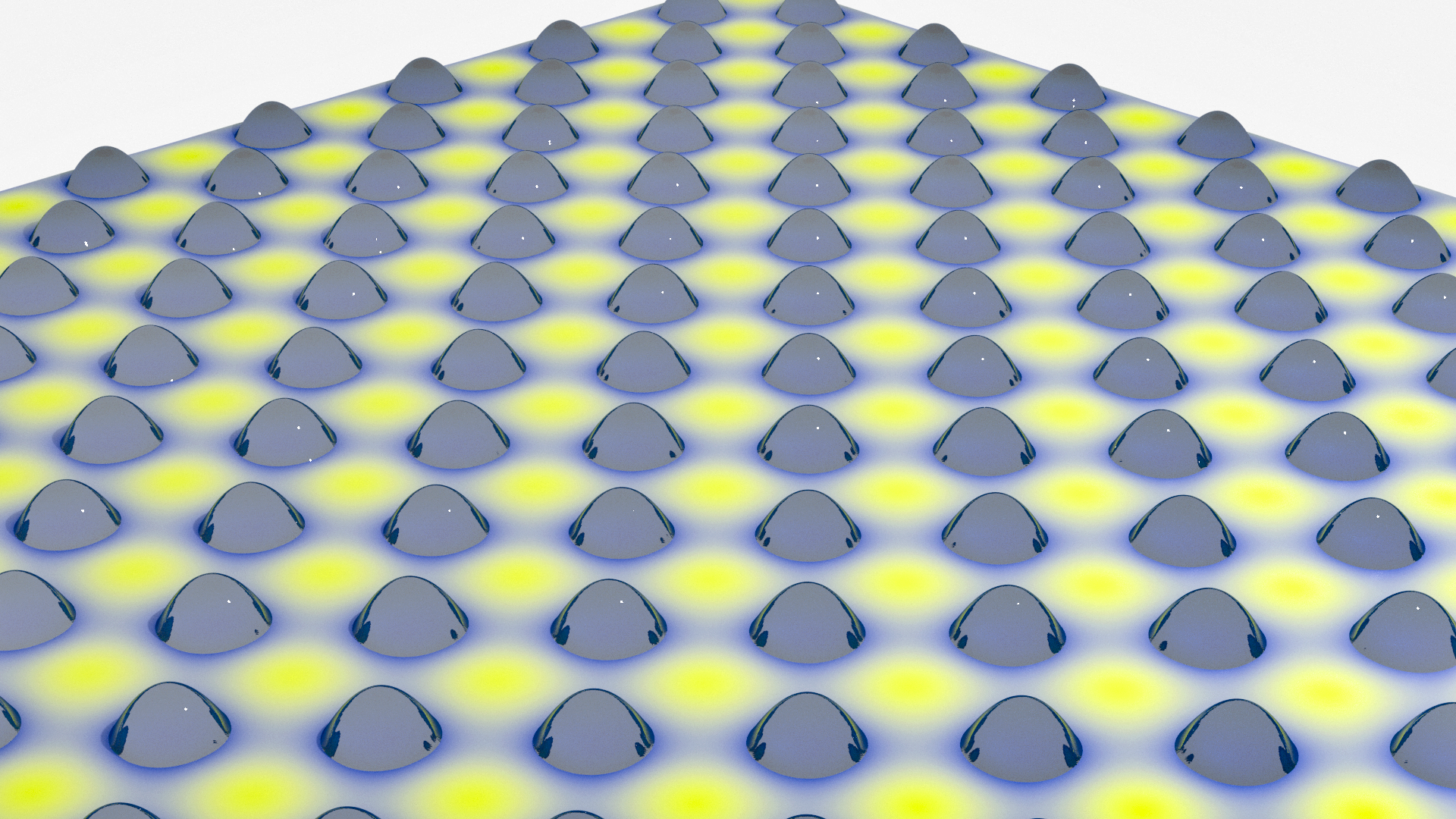}
    \caption{Stationary film thickness field ($t>20t_0$) showing the formation of droplets. The color map indicates the contact angle pattern 
    (Eq.~(\ref{eq:sinetheta}) with $v_{\theta}=0$), with lower (higher) values coded in light blue (yellow).
    }
    \label{fig:handtheta}
\end{figure}
All our simulations are run on a bi-periodic square domain of size $L \times L$ with $L = 512$.
To regularize the contact line divergence~\cite{huh1971hydrodynamic} we use a precursor layer thickness $h_{\ast}=0.07$ and a slip length $\delta = 1$ (see Eqs.~(\ref{eq:mobility})-(\ref{eq:disjoinpressure})). 
The liquid film is initialized with a height field slightly perturbed around the mean valued $h_0=1$, i.e.
\begin{equation}\label{eq:hinitial}
    h(\mathbf{x},0) = h_0 \left[1 + 0.1 \left(\sin\left(\frac{2\pi x}{L}\right)\sin\left(\frac{2\pi y}{L}\right)\right)\right].
\end{equation}
Various wavelengths, in the range $\lambda \in [L/9, L]$, and velocities, $v_{\theta} \in [0.1, 10]v_0$, are considered for the wettability pattern, Eq.~(\ref{eq:sinetheta}). 
In Fig.~\ref{fig:handtheta} we show $h(\mathbf{x},t)$ (droplets) and $\theta(\mathbf{x})$ (color coded) for $v_{\theta} = 0$ (i.e. the static case) and $\lambda = L/2$~\footnote{For a better visualization we take twice the domain length $L$ and periodically continue the image.}, in late stages of dewetting.
As expected, droplets form in regions of small contact angles (blue) while the regions of high contact angles (yellow) dewet.\\

\begin{figure}
    \centering
    \includegraphics[width=0.45\textwidth]{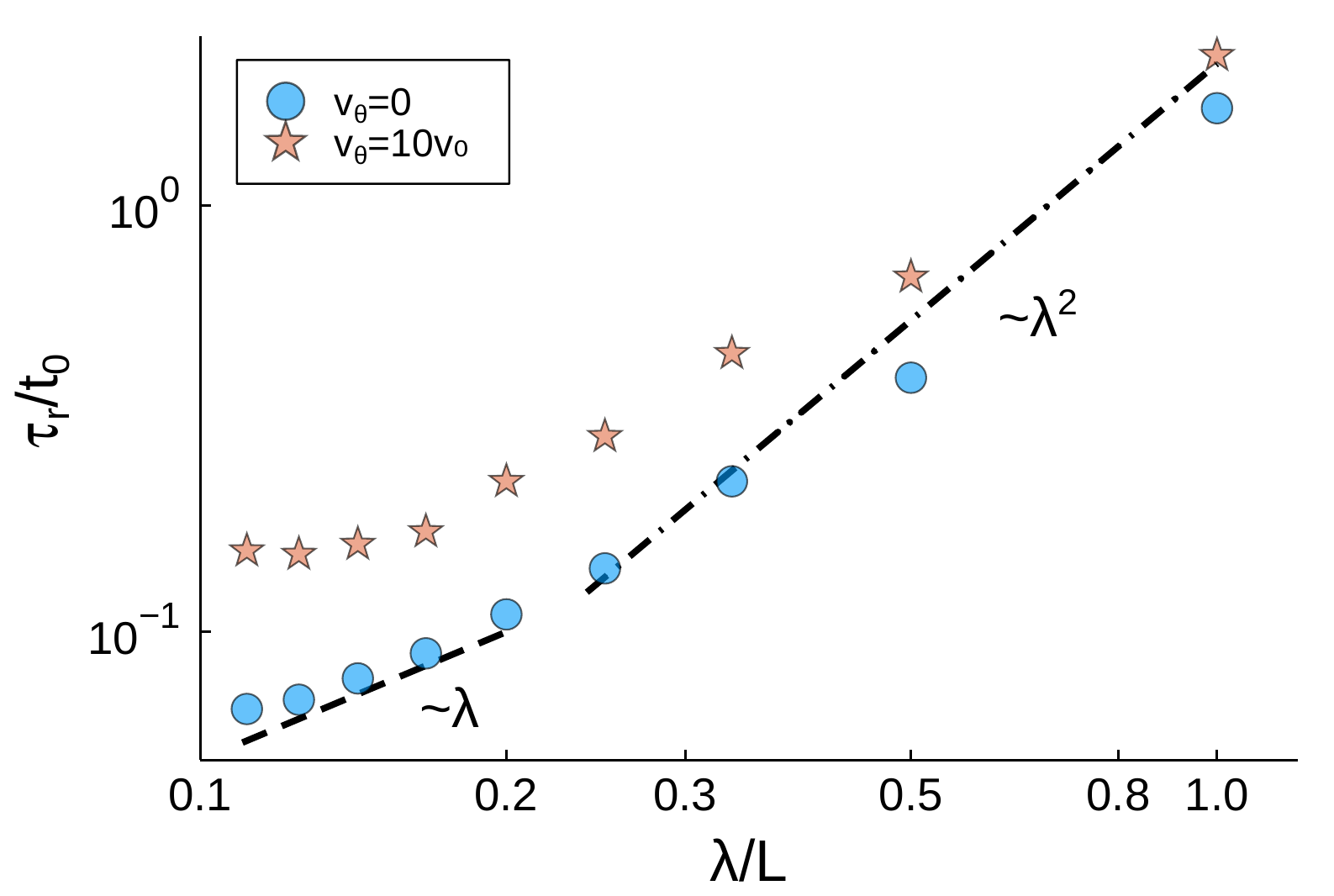}
    \caption{Rupture times $\tau_r$, normalized by $t_0$, as a function of the pattern wavelength $\lambda$, for $v_{\theta}=0$ (\textcolor{jlblue}{$\bullet$}) and $v_{\theta}=10 v_0$ 
    (\textcolor{jlorange}{$\star$}).
    The continuous and dashed lines indicate the linear, $\sim \lambda$, and quadratic, $\sim \lambda^2$, scaling laws, respectively.
        }
    \label{fig:model_rt}
\end{figure}

\noindent {\it Results.} We start our analysis studying, first, how the rupture times depend on the parameters characterising the wettability pattern, namely the wavelength of the contact angle variation, $\lambda$, and wave speed, $v_{\theta}$.
The film rupture time, $\tau_r$, is defined as the least $t$ such that $h(\mathbf{x},\tau_r)=h_{\ast}$ (that is, when the free surface "touches" the substrate).
In Fig.~\ref{fig:model_rt} we report the rupture times, as a function of the wavelength, for stationary ($v_{\theta}=0$) and time-dependent ($v_{\theta}=10 v_0$) patterns. 
It is conveyed that, overall, rupture occurs earlier on the static substrate, suggesting that the time variability tends to stabilize the film, in agreement with linear stability analysis results~\cite{suman2006dynamics}.
In the stationary case we observe that $\tau_r$ grows linearly with $\lambda$, for short wavelengths, and quadratically for longer $\lambda$.
Also for $v_{\theta}=10 v_0$ the rupture times tend to approach the $\lambda^2$ scaling at large $\lambda$, whereas they tend to saturate to a constant value for short wavelengths.
These facts can be explained, qualitatively, as follows. 
Let us first notice that the film rupture consists of two processes, each with its associated time scale: the growth of unstable interface perturbation with rate $t_{\theta}^{-1}$, and the retraction of liquid from more hydrophobic regions, occurring in a characteristic time $t_R$. For longer pattern wavelengths, the dewetting instability amplifies more slowly, such that it is reasonable to take it as the process determining the rupture time, $\tau_r \sim t_{\theta}$.
In this case, from the linearized thin-film equation (in one spatial dimension, for simplicity), obtained setting $h=h_0 + \delta h$ with $\delta h \ll h_0$, 
we can easily see that the exponential growth of the height perturbation is affected by the wettability pattern (variable contact angle) in such a way that $\partial_t (\delta h) \propto (\partial_x^2 (\partial_h\Pi(h_0))) \delta h$. Therefore, since the characteristic time $t_{\theta}$ can be estimated dimensionally as $t_{\theta} \sim \delta h/\dot{(\delta h)}$, the rupture time should go as
\begin{multline}\label{eq:taur_l2}
    \tau_r \sim t_{\theta} \sim  \delta h/\dot{(\delta h)} \propto \frac{3\mu}{h_0^3}(\partial_x^2 (\partial_h\Pi (h_0)))^{-1} \sim \\
    t_0 \left(\frac{q_{\theta}}{q_0}\right)^{-2} \propto t_0 q_0^2 \lambda^2.
\end{multline}
Conversely, for fast growths ($t_{\theta} \ll t_R$), retraction dominates and fixes the time scale, $\tau_r \sim t_R$. 
The latter is related to the time the liquid takes to flow out of regions of high contact angle, whose size is $\sim \lambda$. Hence we have 
\begin{equation}\label{eq:taur_l1}
 \tau_r \sim \tau_R \propto \frac{1}{U_{\Theta}}\lambda,
\end{equation}
where $U_{\Theta}$ is the retraction speed $U_{\Theta} = \frac{\gamma \Theta^3}{9\mu}$~\cite{Edwardse1600183}, with $\Theta = \max_{\mathbf{x}}\theta(\mathbf{x})$.
Interestingly, this phenomenology is qualitatively supported also by the time-dependent case, $v_{\theta} = 10v_0$ (reported in Fig.~\ref{fig:model_rt} as orange stars).
For small $\lambda$, the rupture times tend to saturate to a constant value because, if the pattern wave velocity is large enough to dominate over $U_{\Theta}$, it sets the retraction speed, such that the characteristic time becomes
\begin{equation}\label{eq:taur_saturation}
 \tau_r \sim \tau_R \sim \frac{\lambda}{v_{\theta}} \propto \frac{\lambda}{v_0} = \lambda\frac{t_0}{\lambda} \sim \mbox{const}.
\end{equation}

We now focus on the long time dynamics and in particular on the characterization of the dewetting morphologies and how they are affected by the speed of the wettability wave.
\begin{figure}
    \centering
    \includegraphics[width=0.45\textwidth]{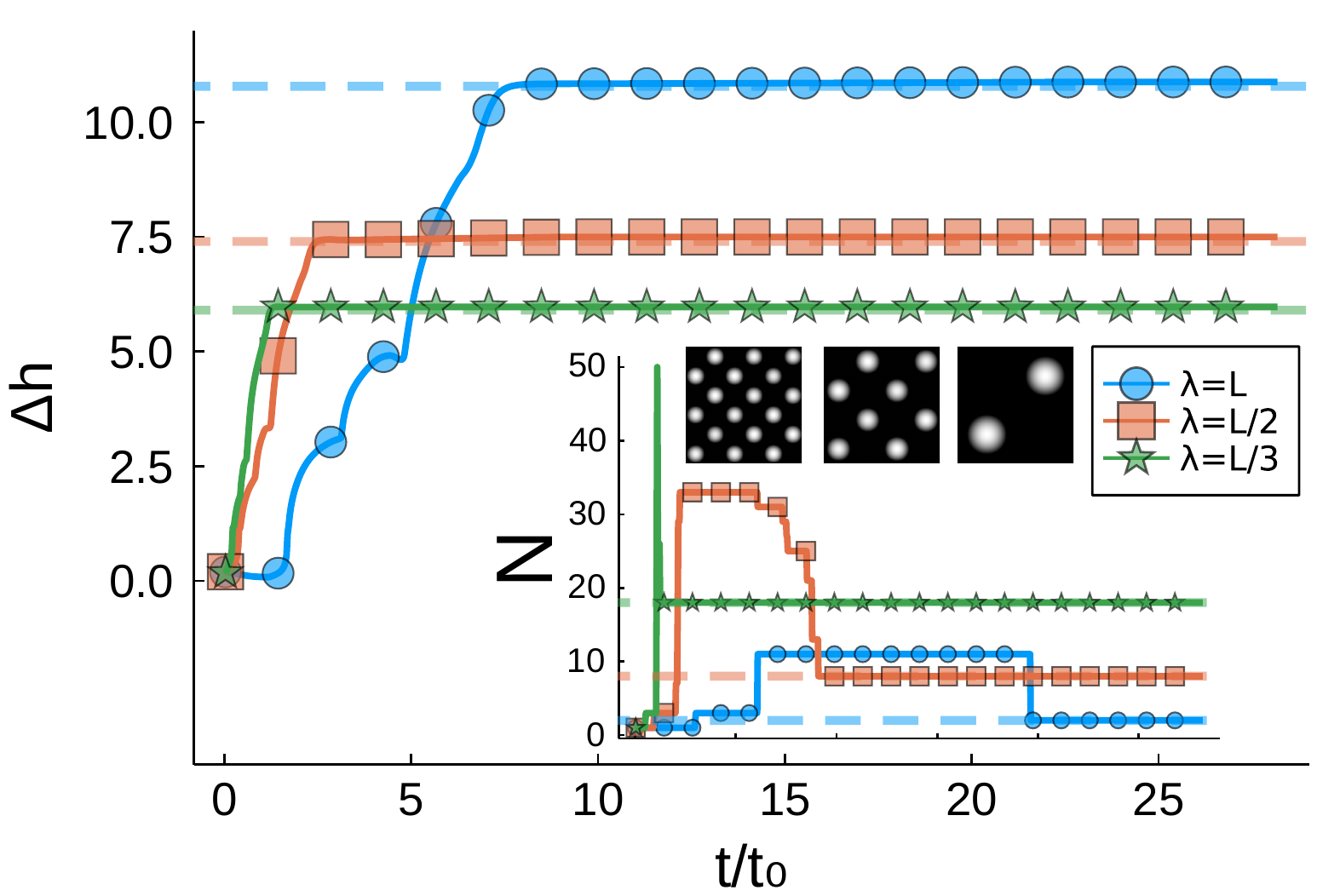}
    \caption{MAIN PANEL: Time evolution of the height fluctuations, $\Delta h(t)$, during the dewetting process on the patterned substrate given by       
    Eq.~(\ref{eq:sinetheta})
    with $v_{\theta}= 0$ and $\lambda=L$ (blue circles), $\lambda=L/2$ (orange squares) and $\lambda=L/3$ (green stars).
    The dashed lines indicate the geometrically expected values of the droplet height $h_d$, assuming monodispersity and a perfect spherical cap shape. 
    INSET: Number of droplets, $N(t)$, as a function of time. The three horizontal dashed lines indicate the number of minima of Eq.~(\ref{eq:sinetheta}),
      which is $2\left(\frac{L}{\lambda}\right)^2$. In the insets, we display snapshots of the stationary droplet states as grey-scale images
      of the film thickness field, $h(\mathbf{x},t)$.
      }
    \label{fig:clusters_v0_sine}
\end{figure}
On the stationary substrate, after rupture all the fluid accumulates in droplets centered at contact angle minima.
Consequently, as we see from the inset of Fig.~\ref{fig:clusters_v0_sine}, where we plot the number of droplets $N(t)$ in time\footnote{A droplet is identified by the set (``cluster'') of points, in the plane, constituting each of the connected components of the set $\{\mathbf{x} \in [0,L]^2 | h(\mathbf{x},t) \geq h_{\ast}$\}; the clusters are determined by means of an algorithm of Hoshen-Kopelman type~\cite{HK}.}, in the steady state ($t \gg t_0$) $N(t)$ attains the value $N_{\infty} = 2(L/\lambda)^2$ (reported as horizontal lines), which equals the minima of Eq.~(\ref{eq:sinetheta}), for $v_{\theta}=0$.
Notice that the number of droplets converges faster for smaller pattern wavelengths, in line with the observation reported and justified in the previous section that the characteristic dewetting time decreases with the wavelength.

As a further consistency check, we follow the evolution of the quantity
\begin{equation}\label{eq:deltah}
  \Delta h(t) = \max_{\mathbf{x}}\{h(\mathbf{x},t)\}-\min_{\mathbf{x}}\{h(\mathbf{x},t)\}.
\end{equation}
In the steady state, since the droplets are essentially monodisperse, this observable represents a measure of the mean droplet height $h_d$. Assuming that the droplet shape is a spherical cap, it can be estimated as $h_d =\left(\frac{3V_d(1-\cos(\theta_d))}{\pi(2+\cos(\theta_d))}\right)^{1/3}$, where $V_d = \frac{h_0 \lambda^2}{2}$ is the droplet volume and $\theta_d$ is the local contact angle (i.e. the value at the contact line). This, in turn, depends on $h_d$, through Eq.~(\ref{eq:sinetheta}), thus making the one above an implicit equation to be solved numerically for $h_d = h_d(\lambda)$. 
$\Delta h(t)$ is reported in Fig.~\ref{fig:clusters_v0_sine} for three different wavelengths, $\lambda = \{L/3,L/2,L \}$, together with the theoretically expected $h_d(\lambda)$ (depicted with lines), showing excellent agreement.\\
\begin{figure}
    \centering
    \includegraphics[width=0.45\textwidth]{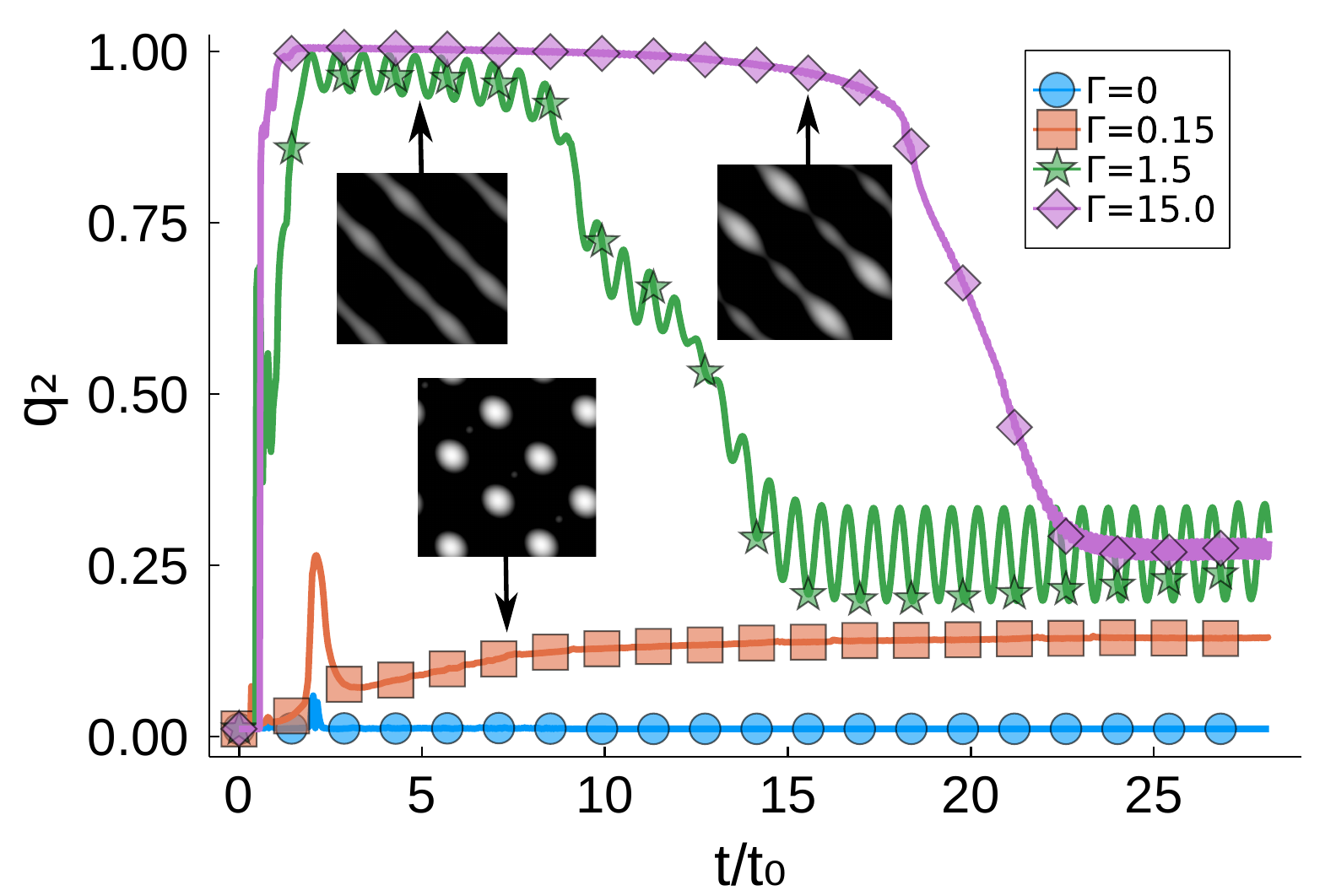}
    \caption{Time evolution of the second order Minkowski structure metric, $q_2(t)$ (see text for details), for different $\Gamma$ values (see Eq.~(\ref{eq:vel_ratio})), on a substrate with pattern wavelength $\lambda=L/2$.
    The grey-scale insets supply snapshots of the film thickness field associated to the various curves.
    }
    \label{fig:msm_q2}
\end{figure}
A time-dependent pattern affects the dewetting morphology quite substantially.
For $v_{\theta} = 0.1 v_0$ we still observe the formation of droplets, similarly to the stationary case ($v_{\theta} = 0$). However, these are transported with the contact angle minima, reproducing a somehow similar behaviour recently described in a numerical study of a droplet on a moving wettability step~\cite{GrawitterStark1}.
If the pattern speed is further increased, for $v_{\theta} = v_0$ we observe the development of rivulet-like structures, aligned with $\mathbf{v}_{\theta}$. 
The film, in fact, while dewetting in the direction normal to the pattern velocity, is exposed, in the direction of the velocity, to a periodic potential with alternating minima and saddle points, which partially (as we will see) stabilizes the film over ``preferential'' lanes along the diagonals. 
This makes the chosen velocity direction, $(1/\sqrt{2},-1/\sqrt{2})$ (or, equivalently, the orthogonal one $(1/\sqrt{2},1/\sqrt{2})$), optimal for the formation of rivulets. 

In order to better characterize the various morphologies we apply the theory of Minkowski's functionals. 
In particular, we employ the second order Minkowski structure metric, $q_2$~\cite{doi:10.1063/1.4774084, Schaller2020}, which can be computed from a Voronoi tessellation of the set of discrete points $(x_i, y_i)$ on the 2D lattice, such that the height field lies above a certain threshold\footnote{The expression is: $q_2 = \frac{1}{N}\sum_j \frac{1}{P_j}|\sum_k L^{(j)}_k e^{2i\phi^{(j)}_k}|$, where the inner sum runs over the edges of length $L^{(j)}_k$, of the $j$-th Voronoi cell, whose perimeter is $P_j$, and $\phi^{(j)}_k$ is the polar angle of the normal to the $k$-th edge. 
The outer sum represents an ensemble average over the $N$ points in the set.}. The $q_2$ metric quantifies the degree of anisotropy of the dewetting morphology, so it takes relatively large values if the structures formed display a preferential direction. 
Measuring $q_2$ then enables us to clearly distinguish between the formation of droplets and rivulets: much larger $q_2$ values are attained for the latter type of structure, as we can see in Fig.~\ref{fig:msm_q2}.
We observe, on the other hand, that such rivulets are metastable and eventually break up into droplets, as indicated by the collapse of $q_2$ at later times. 
Notice, though, that the $q_2$ signal for any $v_{\theta} >0$ always stays above the one for the static case, suggesting that even the smallest pattern velocity introduces a sizeable deformation of the spherical cap shape.
\begin{figure}
    \centering
    \includegraphics[width=0.45\textwidth]{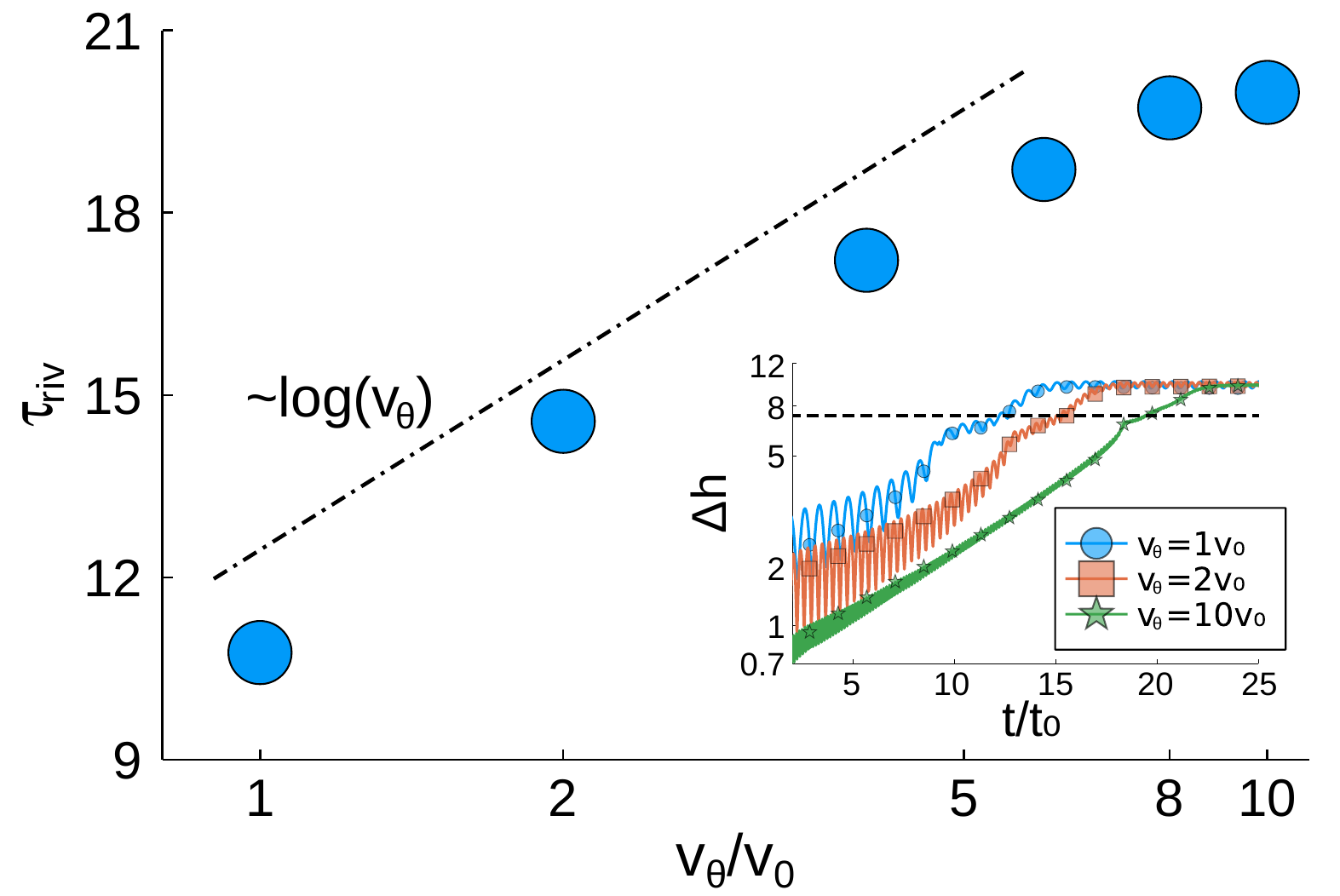}
    \caption{MAIN PANEL: Rivulets life-times, $\tau_{\text{riv}}$, for various pattern velocities.
    The dash-dotted line is plotted as a guide to the eye to highlight the logarithmic dependence, in agreement with 
    the theoretical prediction, Eq.~(\ref{eq:rivlt}).
    INSET: Height fluctuations, $\Delta h(t)$, vs time, computed along the rivulet axis, for three different $v_{\theta}$ values.
    }
    \label{fig:stab_ligs_lam2}
\end{figure}
The breakup is the result of a varicose mode of the rivulet~\cite{doi:10.1063/1.3211248, PhysRevE.77.061605}, whose wavelength is $\approx\lambda$, such that only $N_{\infty}/2$ droplets are counted after breakup. 
These droplets show a peculiar dynamics, characterized by a periodic sequence of spreading and retraction, driven by the pattern, that we dub "pumping state" (see movie file ligament\_formation\_and\_breakup.mp4 in the Supp. Mat.).
The emergence of rivulets is controlled by the competition of two characteristic velocities: the pattern wave speed, $v_{\theta}$, and the retraction speed, $U_{\Theta}$, introduced in Eq.~(\ref{eq:taur_l1}). 
If $U_{\Theta}$ is large as compared to $v_{\theta}$, the film retraction is faster than the transport due to the contact angle field and thus droplets from. However, if $v_{\theta}$ is larger than $U_{\Theta}$, then the retracting film has too little time to form droplets and ends up in the metastable rivulet state.
Recalling the expression (\ref{eq:t0}) for the reference velocity $v_0$, we define the parameter 
\begin{equation}\label{eq:vel_ratio}
    \Gamma = \frac{v_{\theta}}{U_{\Theta}} = \frac{3\lambda h_0^3 q_0^4}{\Theta^3}\chi 
\end{equation}
as the ratio of these two velocities,
where $\chi \equiv v_{\theta}/v_0$. We see from Fig.~\ref{fig:msm_q2} that indeed rivulets form only for $\Gamma > 1$. 
Moreover, the larger $\Gamma$, the more stable the rivulets are; in other words, the rivulet life-time, $\tau_{\text{riv}}$, that can be conventionally taken as the time at which the drop of $q_2$ occurs, grows with $v_{\theta}$ (see Fig.~\ref{fig:stab_ligs_lam2}).
The rivulet itself is, in fact, prone to dewetting, with the liquid accumulating over patches around contact angle minima. However, as the pattern moves, the instability is tamed due to configurations whereby higher contact angle regions underlie height field maxima, thus tending to revert the fluid flow.
Heuristically speaking, this means that, if we define $\Delta h(t)$ as in Eq.~(\ref{eq:deltah}), but restricted on the rivulet axis, this should grow exponentially (with a certain growth rate) only when the system is in the unstable configuration. Namely, $\Delta h(t) \propto e^{\alpha t}$ (see inset of Fig.~\ref{fig:stab_ligs_lam2}) with a prefactor proportional to the time spent by the rivulet in such a configuration, which goes as $\sim \lambda/v_{\theta}$, therefore $\Delta h(t) \sim \alpha (\lambda/v_{\theta})e^{\alpha t}$. 
The rivulet life-time can be seen as the rupture time of the structure along its axis, hence such that $\Delta h (\tau_{\text{riv}}) \sim h_0$ \cite{PhysRevE.104.034801}, which yields
\begin{equation}\label{eq:rivlt}
    \tau_{\text{riv}} \sim \log(v_{\theta}). 
\end{equation}
This logarithmic dependence is indeed observed in the numerical data as shown in Fig.~\ref{fig:stab_ligs_lam2}.

\noindent {\it Conclusions.}
We have presented results, from numerical simulations and theoretical analysis, on the dewetting of thin liquid films on a switchable substrate, modelled with a space and time periodically varying contact angle in the thin-film equation.
Studying how the stability of the film depends on the underlying static pattern, we found that the rupture times grow linearly with the pattern wavelength, for short wavelengths, and quadratically in the long wavelength limit. 
In the time-dependent case, the rupture times are in general longer, indicating an induced greater film stability, and, while the quadratic growth is preserved at long wavelengths, a plateauing behaviour is observed as the wavelength decreases. 
A theoretical explanation has been provided for all these various regimes.
Furthermore, we showed that, at increasing the wettability wave speed, a transition occurs in the dewetting morphology from a multi-droplet to a metastable multi-rivulet state. 
A dimensionless parameter, $\Gamma$, controlling the transition has been identified in the ratio of the pattern speed and the typical film retraction speed over the substrate. 
The rivulets life-time itself grows with the pattern speed, displaying a logarithmic dependence that has been captured by means of phenomenological arguments.
On a broader perspective, our work suggests that switchable substrates offer a new avenue to control thin film dewetting, with obviously relevant implications, for instance, for open microfluidic devices, and paves the way to future studies in this direction, exploiting more complex and dedicated space-time dependencies.

\noindent {\it Acknowledgements}. SZ and JH acknowledge financial support from the Deutsche Forschungsgemeinschaft (DFG) within the priority program SPP2171 ``Dynamic Wetting of Flexible, Adaptive, and Switchable Substrates'', project HA-4382/11.


%
\end{document}